\RequirePackage{fix-cm}
\documentclass[smallextended]{svjour3}       
\smartqed  
\usepackage{graphicx}
\usepackage{amsmath}
\usepackage{amssymb}
\usepackage{graphicx}
\usepackage{cite}
\usepackage{caption}
\usepackage{subcaption}
\usepackage[normalem]{ulem}

\begin{document}

\title{Activity induced nematic order in isotropic liquid crystals
}


\author{Sreejith Santhosh         \and
        Mehrana R. Nejad \and 
        Amin Doostmohammadi \and
        Julia M. Yeomans \and
         Sumesh P. Thampi
}


\institute{Sreejith Santhosh \at
Department of Physics,
Indian Institute of Technology Madras
Chennai - 600036, India
           \and
Mehrana R. Nejad \at
             The Rudolf Peierls Centre for Theoretical Physics, Clarendon Laboratory, Parks Road
Oxford, OX1 3PU, UK
\and
Amin Doostmohammadi
\at
 Niels Bohr Institute, Blegdamsvej 17, 2100, Copenhagen, Denmark
 \and
 Julia M. Yeomans
 \at
 The Rudolf Peierls Centre for Theoretical Physics, Clarendon Laboratory, Parks Road
Oxford, OX1 3PU, UK
\and
 Sumesh P. Thampi \at  Department of Chemical Engineering, 
Indian Institute of Technology Madras
Chennai - 600036, India
}

\date{Received: date / Accepted: date}

\maketitle

\begin{abstract}
We use linear stability analysis to show that an isotropic phase of elongated particles with dipolar flow fields can develop nematic order as a result of their activity. We argue that ordering is favoured if the particles are flow-aligning and is strongest if the wavevector of the order perturbation is neither parallel nor perpendicular to the nematic director. Numerical solutions of the hydrodynamic equations of motion of an active nematic confirm the results. The instability is contrasted to the well-known hydrodynamic instability of an ordered active nematic.
\keywords{Active Nematics \and Hydrodynamic Instability \and Liquid crystals}
\PACS{47.57.Lj \and 83.60.Wc \and 47.63.Gd}
\end{abstract}

\section{Introduction}


Active systems are driven away from thermodynamic equilibrium by continuous energy injection from their constituent elements. Striking examples in nature are intracellular cytoskeletal elements~\cite{sanchez2012spontaneous, hardouin2019reconfigurable, weirich2019self}, cellular tissues \cite{duclos2018spontaneous, saw2017topological}, and bacterial suspensions \cite{di2010bacterial, ramaswamy2010mechanics}, which are all powered by the ability of their individual building blocks - i.e., molecular motors, single cells or individual bacteria - to convert stored or ambient energy into mechanical motion. Similarly, synthetic active particles have been designed to harness energy from the local environment through phoretic mechanisms based on chemical reactions, light or temperature \cite{bechinger2016active, illien2017fuelled}. Consequent generation of mechanical work can combine with microstructural stresses to give chaotic flows and large scale collective patterns of motion in active systems \cite{marchetti2013hydrodynamics, koch2011collective}. 

Nematic order is common in biology and many active materials are well described as active nematics. These include bundles of microtubule filaments and kinesin motors confined to an oil-water interface which generate self assembled yet dynamic structures together with an `active turbulent' flow field \cite{sanchez2012spontaneous, hardouin2019reconfigurable}: chaotic patterns of swirls and jets over length scales larger than the size of individual bundles. An interesting aspect of this active pattern formation is the presence of topological defects of charge $\pm {1}/{2}$, which  is a consequence of the underlying nematic symmetry in the orientation of microtubule filaments \cite{thampi2014vorticity,giomi2013defect,tjhung2012spontaneous}. Existence of short-range nematic order is also important in realizing rather unusual flow states of vortex lattices \cite{shendruk2017dancing}, turbulent to coherent flow transitions in channels \cite{wu2017transition} and in making shape changing active vesicles \cite{keber2014topology, PhysRevLett.123.208001}. Similarly nematic order has also been observed in cellular layers \cite{saw2017topological, Kawaguchi17, duclos2018spontaneous} and in bacterial suspensions~\cite{Volfson08,nishiguchi2017long}. Unlike microtubules and bacteria, shape anisotropy in cells is not obvious, but a numerical study using a phase field approach showed that intercellular stresses elongate isotropic cells thus allowing nematic order to develop \cite{mueller2019emergence}. 

In equilibrium systems a nematic phase is characterised by long-range orientational order, but no long-range positional order. By contrast, orientational ordering is unstable to a hydrodynamic instability  in active systems resulting in active turbulence, characterised by short-range nematic order and flows driven by gradients in the nematic order parameter \cite{simha2002hydrodynamic, voituriez2005spontaneous}. The recent literature deals extensively with this hydrodynamic instability and the consequent stages of evolution, but much less attention has been devoted to  considering the origin of nematic order in active systems in the first place, either experimentally or theoretically. Theoretical models often assume the presence of nematic order by prescribing a free energy, for example a  Landau-de Gennes expansion in terms of the nematic order parameter with parameters chosen to correspond to  the nematic phase. However defining a thermodynamic free energy in non-equilibrium systems is contentious, particularly so when many active systems do not retain nematic ordering as the magnitude of the activity tends to zero. 

Therefore here we provide an alternative, rather natural, mechanism for the generation of nematic order in active systems. We show that activity itself can give rise to nematic order, and that this physics is already present in the dynamical equations of active nematic fluids. In our earlier work \cite{thampi2015intrinsic} we demonstrated the growth of activity induced nematic order numerically and showed that an approximate active - viscous force balance is mathematically equilvalent to an intrinsic free energy in active nematics. Building upon this idea, we now  rigorously show the development of activity induced nematic order in active nematics  through a linear stability analysis, thus predicting the conditions that lead to this order.

We first describe the continuum equations. We then show that active liquid crystals in the isotropic phase are linearly unstable to the development of nematic order and obtain the corresponding dispersion relation. We compare the results from the theoretical analysis to numerical simulations and discuss implications of the new instability with respect to the well-established hydrodynamic instability of an ordered active nematic~\cite{simha2002hydrodynamic}.

\section{Governing Equations}
We consider a continuum model of a suspension of active particles with nematic symmetry. The nematic order of the particles is described by defining a symmetric, traceless, second rank tensor $\mathbf{Q} (\mathbf{x},t)$ \cite{de1993physics}. For uniaxial nematics, $\mathbf{Q} = S(\mathbf{n} \mathbf{n} - \frac{1}{2}\mathbf{I})$ where $S$ gives the magnitude of the order parameter, $\mathbf{n}$ is the director field and $\mathbf{I}$ the identity tensor.

The active nematic equations of motion, a modification of the well known nemato-hydrodynamic equations \cite{beris1994thermodynamics, marchetti2013hydrodynamics}, describe the coupled evolution of the order parameter $\mathbf{Q}$ and the associated incompressible fluid velocity $\mathbf{u}$, 
\begin{align}
    \rho\partial_t u_i + u_k\partial_k u_i =  \partial_j \Pi_{ij}  , \quad \partial_i u_i=0,\label{eqn:ns}\\
    \partial_t Q_{ij} + u_k \partial_k Q_{ij}-\mathcal{W}_{ij}=\Gamma H_{ij},\label{eqn:qevoln}
\end{align}
where the Einstein summation convention over indices is assumed.

The evolution equations for the velocity field, Eq.~(\ref{eqn:ns}), represents momentum and mass conservations with $\rho$ denoting the density. The stress tensor $\boldsymbol{\Pi}$ is composed of three parts,
\begin{equation}
    \boldsymbol{\Pi}=\boldsymbol{\Pi}^{\text{viscous}}+\boldsymbol{\Pi}^{\text{passive}}+\boldsymbol{\Pi}^{\text{active}}.
\end{equation}
The first is the viscous stress, the dissipative contribution due to viscosity, written as a Newtonian constitutive relation \cite{chaikin2000principles},
\begin{equation}
    \Pi_{ij}^{\text{viscous}}=2\eta E_{ij},
\end{equation}
where $\eta$ is the viscosity of the active fluid.

The second contribution is the elastic stress, that generates a back flow due to heterogeneity in the relaxation of the orientational order, \cite{toth2002hydrodynamics}
\begin{align}
\Pi_{ij}^{\text{passive}} &= -P\delta_{ij} + 2 \lambda \left(Q_{ij}+\frac{\delta_{ij}}{2} \right) Q_{kl}H_{lk} \nonumber \\-&\lambda  H_{ik} \left(Q_{kj}+\frac{\delta_{kj}}{2}\right)-\lambda\left(Q_{ik}+\frac{\delta_{ik}}{2}\right)H_{kj} \nonumber\\ &- \kappa (\partial_i Q_{kl})(\partial_j Q_{kl}) + 
    Q_{ik}H_{kj}-H_{ik}Q_{kj},
\end{align}
where $P$ is the pressure field and $\mathbf{H}$ is the molecular field, defined below. The final contribution is the active stress, generated by the microscopic energy conversion mechanism 
\cite{simha2002hydrodynamic, marchetti2013hydrodynamics}, 
\begin{equation}
    \Pi_{ij}^{\text{active}}=-\zeta Q_{ij},\label{eq:active}
\end{equation}
where $\zeta$ is the strength of the activity. The sign of $\zeta$ determines the nature of the stress generated: $\zeta>0$ for extensile stress and $\zeta<0$ for contractile stress. 

The velocity field generated by active stresses is then coupled to the dynamics of the nematic tensor through Eq.~(\ref{eqn:qevoln}) where the generalised advection term,
\begin{eqnarray}
\mathcal{W}_{ij}&=&(\lambda E_{ik}+\Omega_{ik})(Q_{kj}+\frac{\delta_{kj}}{2}) \nonumber \\
&&\!\!\!\!\!\!\!\!\!\!\!\!\!\!\!\!\!\!\!\!\!\!\!\!\!\!\!\!\!\!\!\!
+(Q_{ik}+\frac{\delta_{ik}}{2})(\lambda E_{kj}-\Omega_{kj}) -2\lambda(Q_{ij}+\frac{\delta_{ij}}{2})(Q_{kl}\partial_k u_{l}),
\end{eqnarray}
represents the response of the order parameter $\mathbf{Q}$ to the underlying velocity gradients of the flow field. The symmetric and antisymmetric part of the velocity gradient tensor, namely the strain rate and vorticity tensors, are represented as $\mathbf{E}$ and $\boldsymbol{\Omega}$. Depending upon the value of the tumbling parameter, $\lambda$, active particles may align or tumble in a simple shear flow \cite{edwards2009spontaneous}.
The rotational diffusivity $\Gamma$ determines the time scale of relaxation, driven by a molecular field $\mathbf{H} = \kappa \nabla^2 \mathbf{Q}$ of strength  $\kappa$ which tends to smooth out any gradients in $\mathbf{Q}$. For ordered liquid crystals, $\kappa$ is the orientational elasticity \cite{de1993physics}. 
Note that an 
important distinction in this work is that, unlike most previous studies, we do not prescribe any thermodynamic ordering through the molecular field. Therefore, development of orientational order can only happen through coupling to the velocity field that is produced due to the activity of the particles.
\begin{figure}
\centering
  \includegraphics[width=0.9\linewidth]{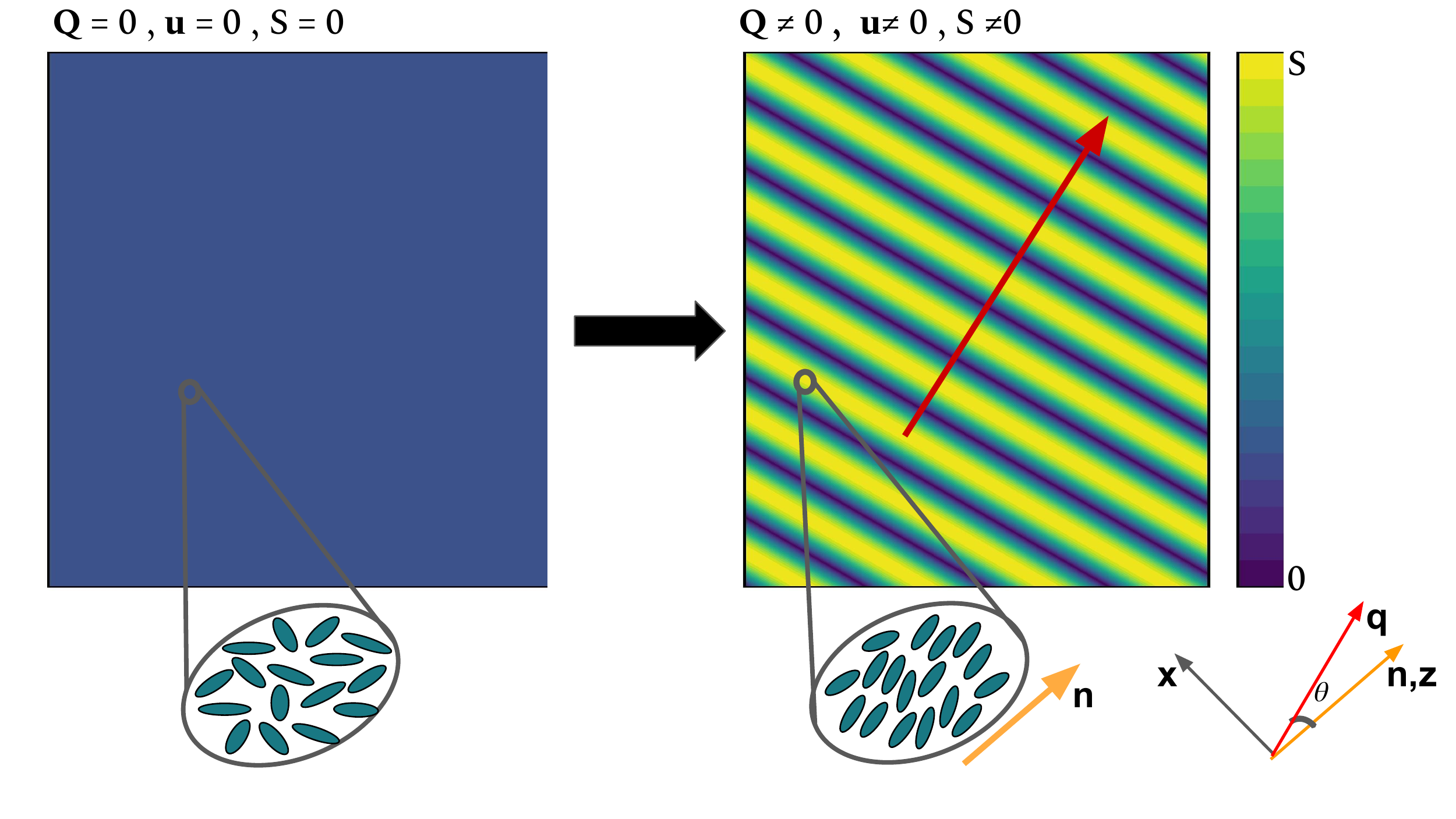}
  \caption{Starting from an isotropic configuration of active particles in a given domain ($S(x,z) = 0$), we introduce a sinusoidal perturbation in $S(x,z)$. The corresponding nematic director field $\mathbf{n}$ is assumed to be uniform in the domain and the wave vector $\mathbf{q}$ makes an angle $\theta$ with $\mathbf{n}$. The coordinate system is chosen such that the $z$-axis coincides with $\mathbf{n}$.}
  \label{fig:basetopert}
\end{figure}

\section{Linear Stability Analysis}

Consider a static fluid, $\mathbf{u} = \mathbf{0}$, of active particles in an isotropic configuration. 
This state corresponds to $S=0$, the state of no orientational order. Superposing a disturbance field $\mathbf{Q}^{\prime}(\mathbf{x},t)$ in the orientational order and linearising Eqs.~(\ref{eqn:qevoln}) and (\ref{eqn:ns}), we obtain the governing equations that determine the evolution of the perturbed order parameter and flow field as
\begin{align}
\partial_t Q^{\prime}_{ij} &= \lambda E^{\prime}_{ij} +\Gamma \kappa \nabla^2 Q^{\prime}_{ij}\label{eqn:Qper}\\
\rho \partial_t u^{\prime}_i &= \partial_j \left( 2\eta E^{\prime}_{ij}-P^{\prime} \delta_{ij}-\lambda \kappa \nabla^2 Q^{\prime}_{ij}-\zeta Q^{\prime}_{ij}\right)\label{eqn:Mper}\\ \partial_i u^{\prime}_i &= 0\label{eqn:Dper}
\end{align}
where perturbed variables have a superscript $^{\prime}$. 


We restrict the perturbation to the magnitude of the order parameter $S^{\prime}$ while assuming that the corresponding director field $\mathbf{n}$ is spatially uniform in the domain as shown in Fig.~\ref{fig:basetopert}. The $z$-axis is chosen along the director field and the wave vector $\mathbf{q}$ makes an angle $\theta$ with the $z$-axis. Introducing the Fourier transform for any fluctuating field $f^\prime$ as $f^\prime(\textbf{r},t)=\int d\omega\: d \textbf{q} \:  \tilde{f}(\textbf{q},t)\: e^{i \textbf{q} \cdot \textbf{r}+\omega t}$, we find the perturbed order parameter $\tilde{S}$ from Eqs.~(\ref{eqn:Qper}) and use incompressibility condition to find $\tilde{\textbf{u}}(\textbf{q},\omega)$ from Eqs.~(\ref{eqn:Mper}):
\begin{align}
\tilde{S} &=\frac{-2 i \lambda q \sin\theta }{\omega +\Gamma \kappa q^2} \tilde{u}_x,\label{eqn:FT1}\\
\tilde{u}_x &=\frac{i \tilde{S} q \sin\theta}{2(\omega \rho +\eta q^2)} \bigg((\zeta-\lambda \kappa q^2)(1+\cos2\theta)\bigg),\label{eqn:FT2}\\
\tilde{u}_z &=\frac{-i \tilde{S} q \cos\theta}{2(\omega \rho +\eta q^2)} \bigg((\zeta-\lambda \kappa q^2)(1-\cos2\theta)\bigg)\label{eqn:FT3}.
\end{align}


%
After algebraic manipulations of Eqs.~(\ref{eqn:FT1})--(\ref{eqn:FT3}) to eliminate Fourier amplitudes we obtain the dispersion relation 
\begin{equation}\label{eqn:disp}
    \rho \omega^2 +q^2(\omega \beta+2\lambda  \cos^2\theta \sin^2\theta (\lambda \kappa q^2-\zeta)+\Gamma \kappa q^2 \eta)=0.
\end{equation}
where $\beta=\eta +\rho \Gamma \kappa $. 
Equation ~(\ref{eqn:disp}) describes the evolution of the perturbed state through a nonlinear relation between the wavenumber $q$ and the frequency $\omega$. It can be seen that activity $\zeta$ plays a destabilising role $\propto q$ in the evolution of the frequency $\omega$, while the stabilising effect of the orientational elasticity, $\kappa$  $\propto q^{2}$. This indicates the existence of a cutoff length $\propto \sqrt{\kappa/\zeta}$, which is normally introduced as the active length scale~\cite{Giomi15, Guillamat2017, Martinez19}. 

\begin{figure}
\centering
  \includegraphics[width=0.7\linewidth]{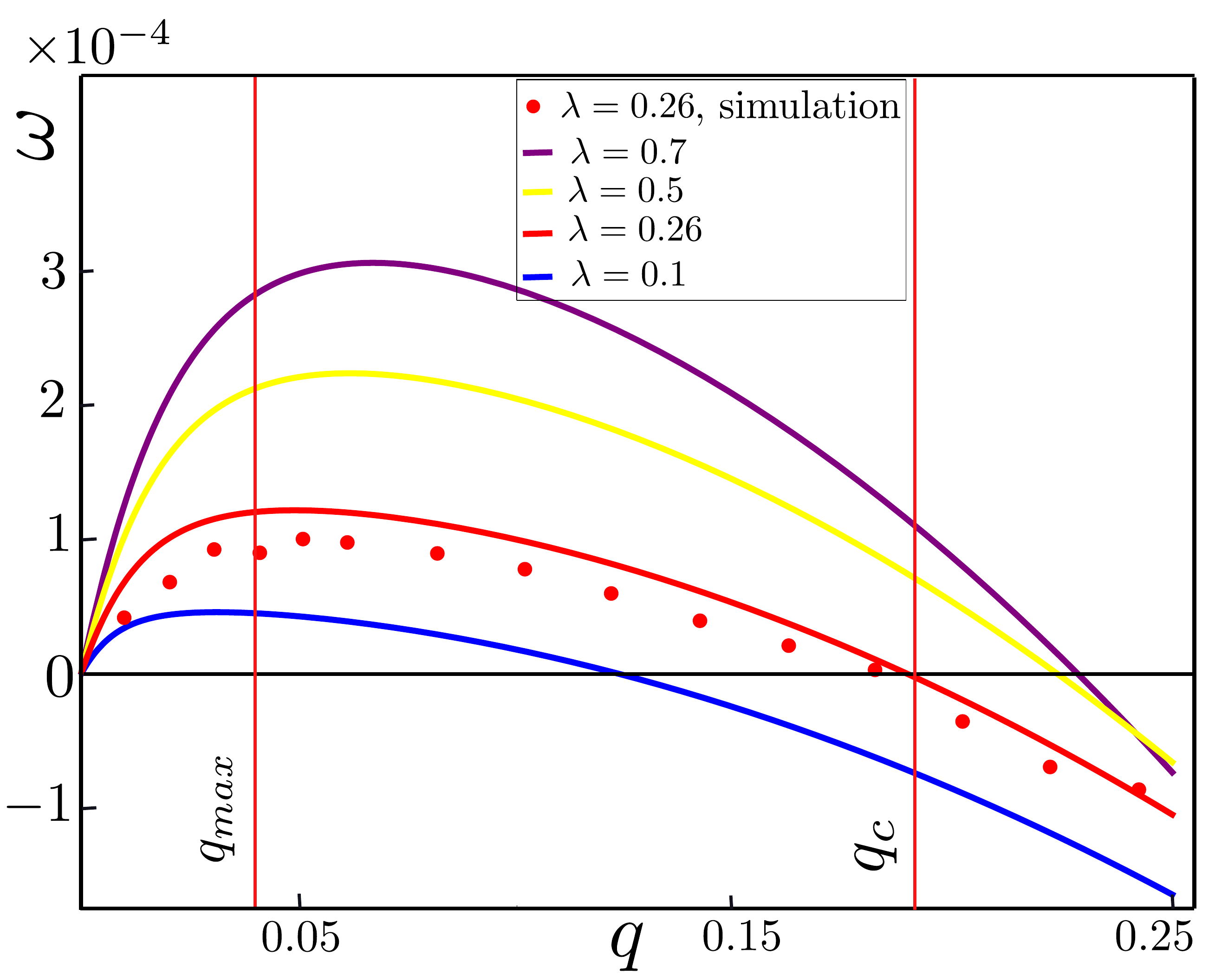}
  \caption{Dispersion relation described by Eq.~(\ref{eqn:disp}) plotted for $\theta = \pi/4$ for different values of the tumbling parameter $\lambda$. The data points obtained for  $\lambda = 0.4$ from numerical simulations are also shown, together with
  $q_{\text{critical}}$ and $q_{\text{max}}$ for this value of $\lambda$. (See section on Numerical Simulations for the other parameters used.) }
  \label{fig:baasdasetopert}
\end{figure}

The system is only unstable to long wavelength perturbations when $\lambda \zeta > 0$. 
 The critical wave number below which the instability occurs, 
\begin{equation}
    q_c=\sqrt{\frac{\lambda \zeta}{2 \kappa(2\eta \Gamma+\lambda^2)}},
     \label{eqn:qcritical}
\end{equation}
is determined from the active length scale $\sqrt{\kappa/\zeta}$, tumbling parameter $\lambda$, and the ratio of viscosities $\eta\Gamma$, where $\Gamma=1/\gamma$ with $\gamma$ the rotational viscosity of the nematic director.

It is instructive to contrast this instability to the well-known hydrodynamic instability of an ordered nematic. As described by Simha \& Ramaswamy~\cite{simha2002hydrodynamic}, the active nematic phase is unstable to bend or splay deformations for any non-zero activity. The linear stability analysis conducted here shows that in the absence of nematic order, there is a critical wave number below which the isotropic configuration of active particles is unstable to the development of nematic order. Once the nematic order is established, the bend or splay hydrodynamic instability can occur.

To further investigate the stability of the isotropic state to small perturbations in the magnitude of the nematic order, we next examine the dispersion relationship for representative sets of physical parameters and angle $\theta$ between the wavevector and the nematic director. Since previous numerical studies have shown that the emergence of nematic order is enhanced by shear alignment of the nematogens \cite{olmsted1992isotropic} we first plot the dispersion relation for various values of the tumbling parameter $\lambda$ (Fig.~\ref{fig:baasdasetopert}).
As evident from Fig.~\ref{fig:baasdasetopert}, $\omega > 0$ for a range of $|\mathbf{q}|$, predicting an exponential growth of the perturbed variables and indicating that the system is linearly unstable. Furthermore, the most unstable wavenumber $q_{\text{max}}$ (corresponding to the fastest growth), increases with increasing tumbling parameter indicating that stronger shear alignment drives smaller wavelengths unstable. An analytical expression for $q_{\text{max}}$ can be obtained as
\begin{align}
    q_{\text{max}}&=\sqrt{\frac{2\lambda\zeta \cos^2\theta \sin^2 \theta}{\frac{2}{3}\kappa J^2+J(\eta+\Gamma\kappa\rho)\sqrt{\frac{\kappa}{3\rho}}}},\\
    J&=\sqrt{3 \eta \Gamma +6 \lambda^2 \cos^2\theta \sin^2\theta}.
    \label{eqn:qmax}
\end{align}

Again it is apparent that the competition between the activity 
and the nematic elasticity is a factor in determining the most unstable modes. 


\begin{figure}
\centering
  \includegraphics[width=0.5\linewidth]{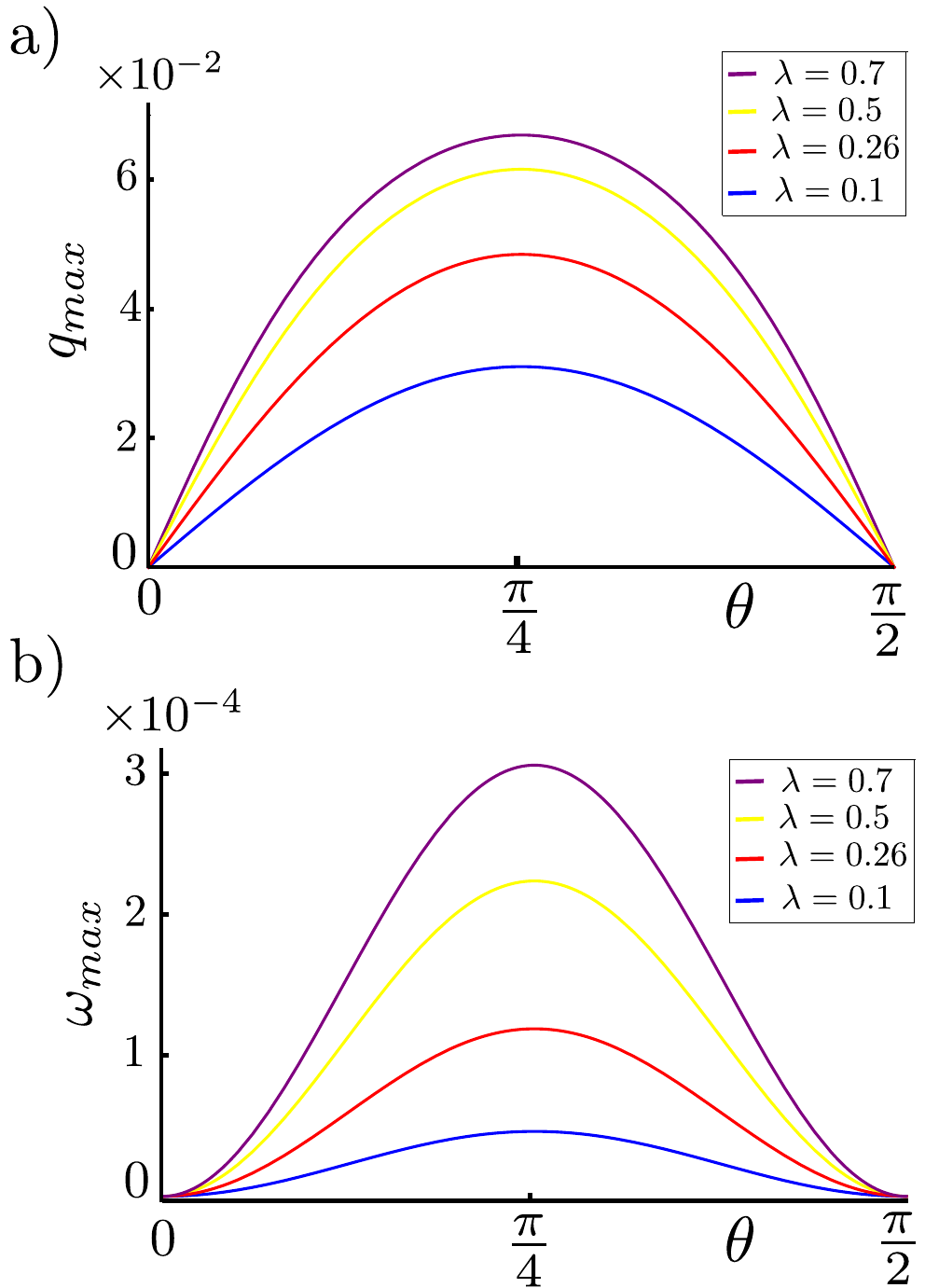}
    \caption{Dependence of (a) the most unstable wavenumber $q_{\text{max}}$ and (b) the  growth rate $\omega_{\text{max}}$ on the angle $\theta$ between the wavevector and the nematic director for various values of the tumbling parameter $\lambda$.}
        \label{fig:qomegamax}
\end{figure} 

Figure~\ref{fig:qomegamax} shows the variation of the most unstable wavenumber $q_{\text{max}}$ and the corresponding growth rate $\omega_{\text{max}}$ as a function of the angle between the wavevector and the director field $\theta$. When $\theta$ approaches $0$ or $\frac{\pi}{2}$  both $q_{\text{max}}$ and $\omega_{\text{max}}$ drop to zero indicating that there is no nematic order development along or perpendicular to the director.
 In order to understand the dependence of the most unstable modes on the alignment parameter and on $\theta$, we next explain the physical mechanism by which activity drives the development of nematic order in an isotropic system.


\section{Physical mechanism}
A conclusion that can be drawn from Fig.~\ref{fig:qomegamax} is that the growth rate of the perturbations in nematic order strongly depends on the tumbling parameter; indeed for $\lambda = 0$ no instability is observed. This indicates that the physical mechanism behind the development of the nematic order should involve the response of active particles to velocity gradients and in particular to the extensional flow field. 

\begin{figure}
    \centering
    \begin{subfigure}[b]{0.75\textwidth}
        \includegraphics[width=\linewidth]{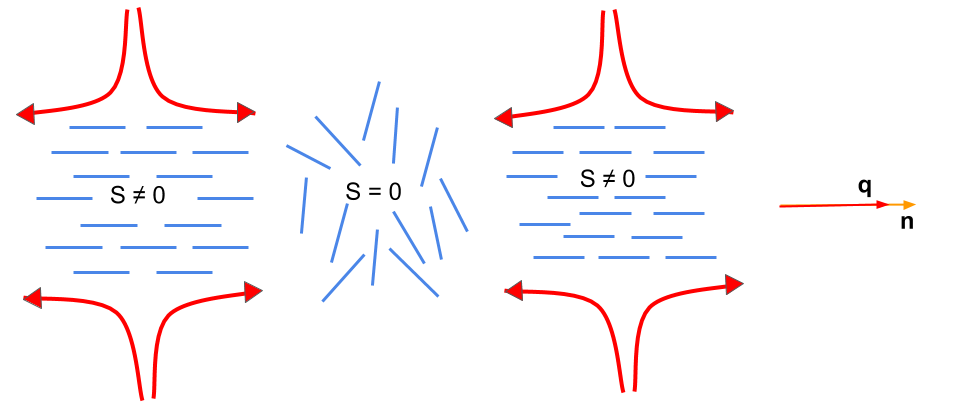}
        \caption{(a) $\theta = 0$}
        \label{a}
    \end{subfigure}\\
    \begin{subfigure}[b]{0.75\textwidth}
        \includegraphics[width=\linewidth]{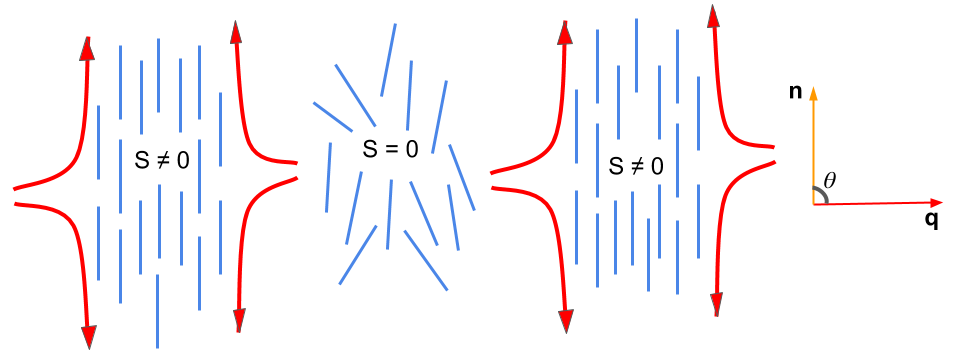}
        \caption{(b) $\theta = \frac{\pi}{2}$}
    \end{subfigure}\\
    \begin{subfigure}[b]{0.75\textwidth}
        \includegraphics[width=\linewidth]{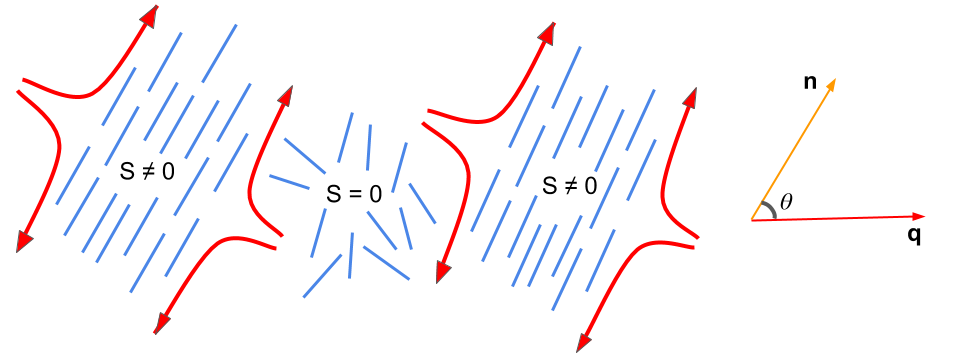}
        \caption{(c) $\theta = \frac{\pi}{4}$}
    \end{subfigure}
        \caption{Physical mechanism for the growth of nematic order in an isotropic system of active particles. Blue solid lines indicate nematic directors and red arrows denote the activity-induced flows. In (c) the disordered region at the centre is aligned by the shear flow set up by the neighbouring ordered regions.}
        \label{fig:physmech}
\end{figure}

A schematic of the physical mechanism involved in the creation of nematic order
is shown in Fig.~\ref{fig:physmech}. Each active particle creates a dipolar flow field. The dipole lies along the long axis of the particle, and extensile particles with $\zeta>0$ pull fluid in from their sides and push it out from their front and back~\cite{Lauga09}. In a perfectly isotropic arrangement of active particles the dipolar flows generated by each particle cancel each other on sufficiently large length scales giving rise to no net flow everywhere in the system. However, if a fluctuation generates a local nematic alignment of active particles, then this results in the generation of locally shearing flow fields. In such shear flows particles align further
in a manner that enhances this shear flow \cite{saintillan2018rheology}. This results in strengthening of the nematic order creating a bootstrap effect: stronger alignment leads to stronger flows that in turn generate stronger alignment. 

The competing tendency of the rotational diffusion is to restore isotropic order of the active particles,
and hence suppress the growth of the instability. Therefore, flow has to established on length scales larger than this orientational diffusion length scale giving rise to a critical wave number for the growth of nematic order. This also explains the dependence of the order development on the tumbling parameter $\lambda$. For larger $\lambda$ the particles more strongly align with the activity-induced flows, therefore enhancing the bootstrap effect.

Furthermore, the dependence on the angle between the fluctuation wave vector and the nematic director can be explained as follows: If the nematic order is developed parallel or perpendicular to the nematic director (Fig.~\ref{fig:physmech} (a), (b)) - corresponding to $\theta = 0,~ \pi/2$, respectively - the resultant extensional flows cancel each other, producing no shear on the isotropic region. As such, any fluctuation decays and both $q_{\text{max}}$ and $\omega_{\text{max}}$ drop to zero, as evident from Fig.~\ref{fig:qomegamax} at $\theta = 0,~ \pi/2$. On the other hand, for any $0 < \theta < \pi/2$, the resultant local activity-induced extensional flows produce a shear flow on the local isotropic regions, inducing further nematic alignment in them. 

\section{Numerical Simulations}
In order to test the predictions from the linear stability analysis we next solve Eqs.~(\ref{eqn:ns})--(\ref{eq:active}) numerically using 
a method of lines for the evolution equation for the order parameter and a lattice Boltzmann method for Navier Stokes equations \cite{marenduzzo2007steady,thampi2014vorticity}. Simulations are performed in a domain of $400 \times 400$ with periodic boundary conditions, and the simulation parameters are chosen in accordance with the values used to plot  the dispersion relations in Fig.~\ref{fig:baasdasetopert}: $\rho = 1, \lambda = 0.26$, $\kappa = 0.01$, $\eta = 0.66$, $\Gamma = 0.34$ and $\zeta=0.0007$ (in lattice Boltzmann units). 

\begin{figure}
     \centering
     \begin{subfigure}[b]{0.49\textwidth}
         \centering
         \includegraphics[width=\linewidth]{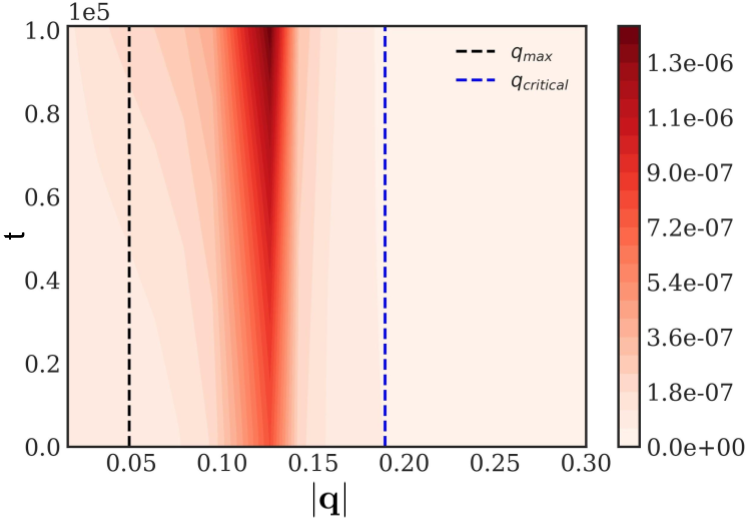}
         \caption{(a)}
         \label{fig:sin_init}
     \end{subfigure}
     \hfill
     \begin{subfigure}[b]{0.49\textwidth}
         \centering
         \includegraphics[width=\linewidth]{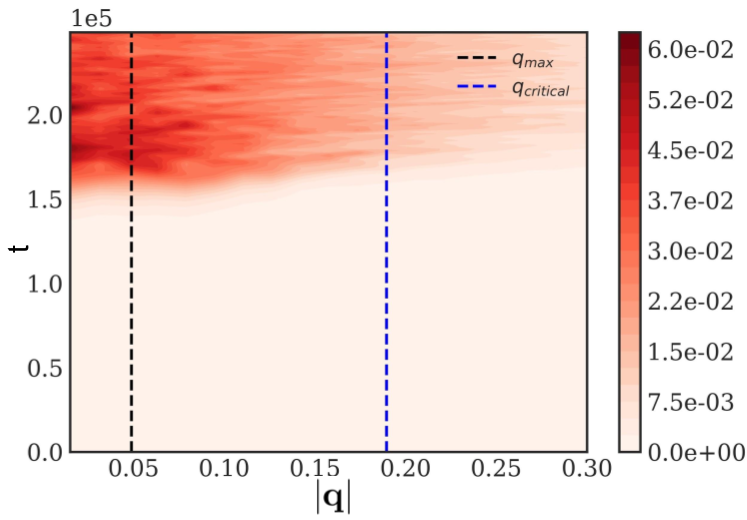}
         \caption{(b)}
    \label{fig:random_init}
     \end{subfigure}
        \label{fig:three graphs}
        \caption{Kymographs of the evolution of the Fourier amplitude of the magnitude of the nematic order, $\hat{S}$, shown by the colormap, as a function of the  wavenumber $|\mathbf{q}|$ (on the horizontal axis) and time t (on the vertical axis) for (a) the sinusoidal initial condition $S = S_0 + B \sin{\mathbf{q}\cdot\mathbf{r}}$ with 
        $S_0 = 10^{-6}$, $B = 10^{-7}$,  $q = 0.12$, $\theta = \frac{\pi}{4}$  
         and $\mathbf{u} = 0$, (b) for a randomly initialised $\mathbf{Q}$; with $S(x,z)$ taking random values from $(0,10^{-7})$ and $\mathbf{n}$ constant along the $x$-axis.
        In (a) the Fourier transform is performed along the perturbation wavevector $\mathbf{q}$ and in (b) the Fourier transform is performed along an arbitrarily chosen axis.}
\end{figure}

In order to emulate the initial perturbation, simulations are started with a well defined initial condition, prescribing the perturbation in the magnitude of the nematic order parameter through a sine function $S = S_0 + B \sin{\mathbf{q}\cdot\mathbf{r}}$ and taking the system to be at rest, $\mathbf{u} = 0$ at time $t=0$.
The corresponding, spatially uniform, director field $\mathbf{n}$ was prescribed to lie at an angle $\theta=\pi/4$ with $\mathbf{q}$. The result obtained from the simulation is presented in Fig.~\ref{fig:sin_init} in the form of a kymograph where the horizontal axis is the wavenumber ($q$-space), the vertical axis is time and the color field indicates the Fourier amplitude of the order parameter, $\hat{S}$.  Thus, Fig.~\ref{fig:sin_init} represents the spatio-temporal evolution of the initial sine wave perturbation in the magnitude of the nematic order. The simulation results confirm the predictions of linear stability analysis, showing that the imposed perturbation grows at the excited wavenumber ($|\mathbf{q}| = 0.12$). Moreover, simulations show that, at later times, larger wavelengths (smaller wavenumbers $|\mathbf{q}| < 0.12$) are excited, an aspect beyond the scope of linear stability analysis. However, the spread towards smaller $|\mathbf{q}|$ is consistent with the long wavelength instability predicted by the dispersion relation.

In order to quantitatively compare the simulation results with the results from linear stability analysis, we quantified the growth rate of the order parameter. The spatially averaged value of $\left\langle({S(x,z)} - S_0)^{2}\right\rangle$ was calculated as a function of time and fitted to an exponential function $e^{2 \omega_s t}$ to extract the growth rate $\omega_s$. The growth rates obtained as a function of $q$ are shown in Fig.~\ref{fig:baasdasetopert} and match well with the analytical results.

Finally we carried out a similar analysis by performing numerical simulations on a system perturbed with a random field $S$. The results are presented as a kymograph in Fig.~\ref{fig:random_init}. At $t=0$, there is no dominant wave. As time proceeds, regions $q < q_{\text{critical}}$ grow predominantly indicating the growth of nematic order in the isotropic phase.
 
 \section{Discussion}
 
%
We have used the continuum equations of active nematics, without the assumption of any underlying thermodynamic ordering, to establish how
activity, which arises from the flow fields of the constituent particles, can give rise to nematic order.  While activity tries to generate nematic order in the system, the rotational diffusion of active particles acts to destroy this order. This competition results in a critical wave number below which exponential growth of nematic order is predicted. 

The analytical calculations assumed a particular choice of perturbation, namely a uniform director field but a sinusoidally varying magnitude of the order parameter. While this choice is made to simplify the analysis, a generic perturbation will include such terms and will lead to instability.

Hydrodynamically mediated growth of nematic order has also been predicted in isotropic swimmer suspensions using both numerical simulations and a linear stability analysis \cite{subramanian2009critical, saintillan2008instabilities, saintillan2008instabilities2}. In analytical calculations, a Fokker-Planck equation describing the evolution of orientation distribution function is used to describe the orientational order and the linear stability analysis predicts a critical concentration required for the onset of collective motion - a result similar to our observations.  Thus, our work using the theory of active liquid crystals is consistent with the fluid mechanical approach to active matter and establishes a link between the two approaches.

It can be seen from the dispersion relation (\ref{eqn:disp}) that there is a bifurcation for the parameter combination $\lambda \zeta$ \cite{thampi2015intrinsic}.
 If $\lambda \zeta < 0$, then there is no value of $q$ for which $\omega > 0$ and the isotropic system remains stable. $\zeta < 0$ corresponds to contractile systems and $\lambda < 0$ corresponds to disc-like active particles. 
The physical mechanism for the bifurcation is that a rod-shaped particle ($\lambda$ \textgreater 0) attains a stable position in an extensional flow when it is aligned along the extensional axis, while a plate-like particle ($\lambda$ \textless 0) 
attains a stable configuration along the compressional axis (with particle orientation defined along the normal to the plate). 

Finally, we are now in a position to give a more detailed comparison of the mechanism for instability detailed here to the hydrodynamic instability of ordered active nematics described by Simha \& Ramaswamy~\cite{simha2002hydrodynamic}. Nematic order is known to be unstable in active fluids \cite{simha2002hydrodynamic, voituriez2005spontaneous} because of the growth of a bend (or splay) deformation of the director field for extensile (or contractile) systems even when the magnitude of the order parameter ($S$) is uniform. Here the instability we describe represents the growth of the nematic order $S$ even when the director field is uniform. Both of these mechanisms are long range instabilities and they may occur simultaneously. However, the angular dependence is different to leading order:  In an ordered suspension, if the wavevector describing the fluctuations in the director field lies at an angle $ \pi/4$ with the director field, then the growth rate is zero. Here the wavevector describes the variation in the magnitude of the orientational order and the  growth rate is large when it makes an angle $\theta = \pi/4$ with the director field. In the relevant biological systems we expect both instability mechanisms to be present, with the instability associated with the nematic order development - described here - preceding the bend-splay instability of the established nematically ordered state. 

\begin{acknowledgements}
We acknowledge Santhan Chandragiri for helpful discussions. A.D was supported by the Novo Nordisk Foundation (grant agreement No. NNF18SA0035142)
\end{acknowledgements}

%
%

\bibliographystyle{spphys}       

\bibliography{prl}


\end{document}